\documentclass[a4paper]{mem}
\usepackage{natbib}\usepackage{txfonts}
\usepackage{balance}
\usepackage{graphicx}
\usepackage[a4paper]{hyperref}
\idline{75}{282}
\begin{document}
\def\teff{$T\rm_{eff }$}
\def\logg{\mbox{log~{\it g}}}
\def\Msun{\mbox{$M_{\odot}$}}
\def\rpro{\mbox{$r$-process}}
\def\spro{\mbox{$s$-process}}
\def\ncap{\mbox{$n$-capture}}

\title{Low metallicity AGB models: H profile in the $^{13}$C-pocket
and the effect on the $s$-process}

   \subtitle{}

\author{
S. \,Bisterzo\inst{1}
\and S. \, Cristallo\inst{2,3}
          }

  \offprints{P. Bonifacio}

\institute{
Dipartimento di Fisica Generale,
Universit\`a di Torino, 10125 (To) Italy
\and
Departamento de Fisica Teorica y del Cosmos,
Universidad de Granada, Campus de Fuentenueva,
18071 Granada, Spain
\and
INAF Osservatorio Astronomico di Collurania,
via M. Maggini, 64100 Teramo, Italy \\
\email{bisterzo@ph.unito.it}
}

\authorrunning{Bisterzo \& Cristallo}

\titlerunning{H profile in the $^{13}$C-pocket and \spro\ effects}

\abstract{
The $^{13}$C($\alpha$, n)$^{16}$O reaction is the major neutron source
in low mass asymptotic giant
branch (AGB) stars, where the main and the strong $s$ process components
are synthesised.
After a third dredge-up (TDU) episode, $^{13}$C burns radiatively in a
thin pocket which forms in the top layers of the He-intershell, by proton
capture on the abundant $^{12}$C.
Therefore, a mixing of a few protons from the H-rich envelope into the
He-rich region is requested.
However, the origin and the efficiency of this mixing episode are still
matter of debate and, consequently, the formation of the $^{13}$C-pocket
represents a significative source of uncertainty affecting AGB models.
We analyse the effects on the nucleosynthesis of the s-elements
caused by the variation of the hydrogen profile in the region where the
$^{13}$C-pocket forms for an AGB model with $M$ = 2 $\Msun$ and [Fe/H] = $-$2.3.
In particular, we concentrate on three isotopes ($^{89}$Y, $^{139}$La and
$^{208}$Pb), chosen as representative of the three $s$-process peaks.
}
\maketitle{}

\section{Introduction}

During their thermally pulsing (TP) phase, low mass asymptotic giant
branch (AGB) stars are the site of the main and the strong
component of the $s$-process, which is the major responsible for
the nucleosynthesis of half the nuclei from Sr to Pb-Bi.
After a limited number of pulses, 
the convective envelope penetrates into
the He-intershell at the quenching of each convective instability,
mixing freshly synthesized $^{4}$He, $^{12}$C, and \spro\
elements to the surface (third dredge-up, TDU).
\\
The major neutron source in low mass AGB stars is the $^{13}$C($\alpha$,
n)$^{16}$O reaction, which burns radiatively during the interpulse
period in a thin region at the top of the He-intershell ($^{13}$C-pocket).
The physical mechanism that allows the formation of the $^{13}$C-pocket
is a debated problem.
A small amount of protons is assumed to
penetrate from the envelope into the He-intershell during TDU episodes
(Iben \& Renzini 1982).
Then, at H reignition, a large amount of $^{13}$C is synthesised in the
top layers of the intershell via the $^{12}$C(p,
$\gamma$)$^{13}$N($\beta$$^+$$\nu$)$^{13}$C nuclear chain.
This $^{13}$C is of primary origin and, therefore,
independent of the metallicity.
During the interpulse, the H-burning shell advances
in mass, compressing and heating the underlying material, and
at $T$ $\sim$ 0.9 $\times$ 10$^8$ K the $^{13}$C($\alpha$,
n)$^{16}$O reaction starts releasing neutrons in radiative
conditions.
Later on, the synthesised \spro\ nuclei are engulfed and diluted in
the next convective region generated by TP.
\\
Different evolutionary and post-process codes have been developed
in the last years to understand the nucleosynthesis in low mass AGB stars
(e.g., Straniero et al. 1995, 2003; Gallino et al. 1998;
Goriely \& Mowlavi 2000;
 Karakas \& Lattanzio 2003, 2007; Campbell \& Lattanzio
2008; Straniero, Gallino \& Cristallo 2006). 
Several mechanisms have been proposed to reproduce the
mixing leading to the $^{13}$C-pocket formation: semi-convection
(Hollowell \& Iben 1988), models including rotation (Langer et al.
1999; Herwig et al. 2003; Siess et al. 2004), gravity waves
(Denissenkov \& Tout 2003), exponential diffusive overshoot at the
borders of all convective zones (Herwig et al. 1997),
opacity-induced overshoot at the base of the convective envelope
(Straniero et al. 2006). A clear answer to the properties of such
mixing has not been reached yet. 
\\
We test here the effects on the nucleosynthesis of the $s$
elements by adopting different H profiles in the
region of the $^{13}$C-pocket forming after the 1$^{st}$ TDU of an
AGB model with initial $M$ = 2 $M_\odot$ and [Fe/H] = $-$2.3.
Comparison between full evolutionary FRANEC (Frascati Raphson-Newton 
Evolutionary Code) models (Cristallo et al. 2009, hereafter C09) 
and FRANEC models coupled with a post-process nucleosynthesis method
(Gallino et al. 1998; Bisterzo et al. 2010) are presented.

\section{Results}

\begin{figure}[t!]
\resizebox{\hsize}{!}{\includegraphics[angle=-90]{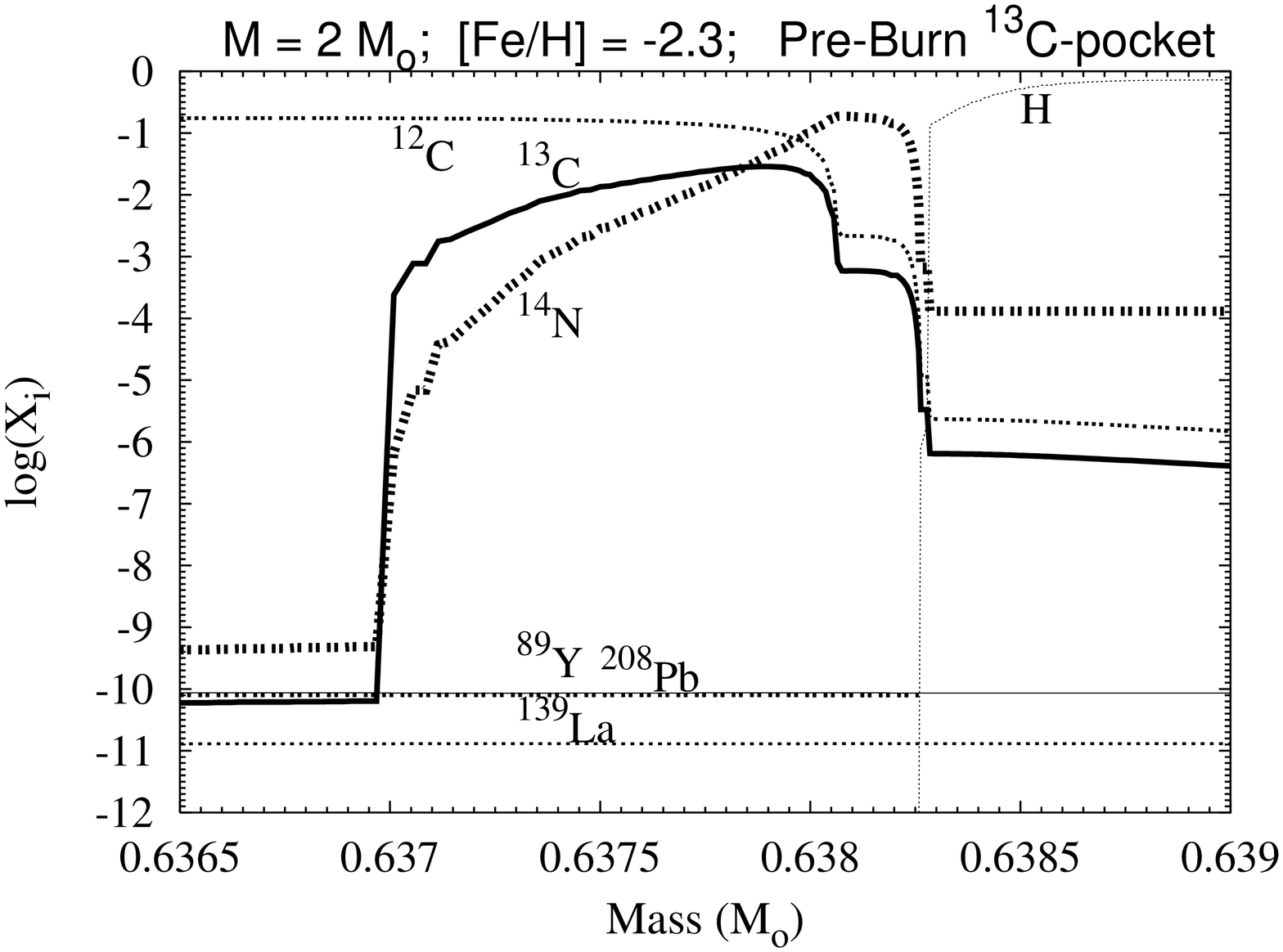}}
\\
\\
\resizebox{\hsize}{!}{\includegraphics[angle=-90]{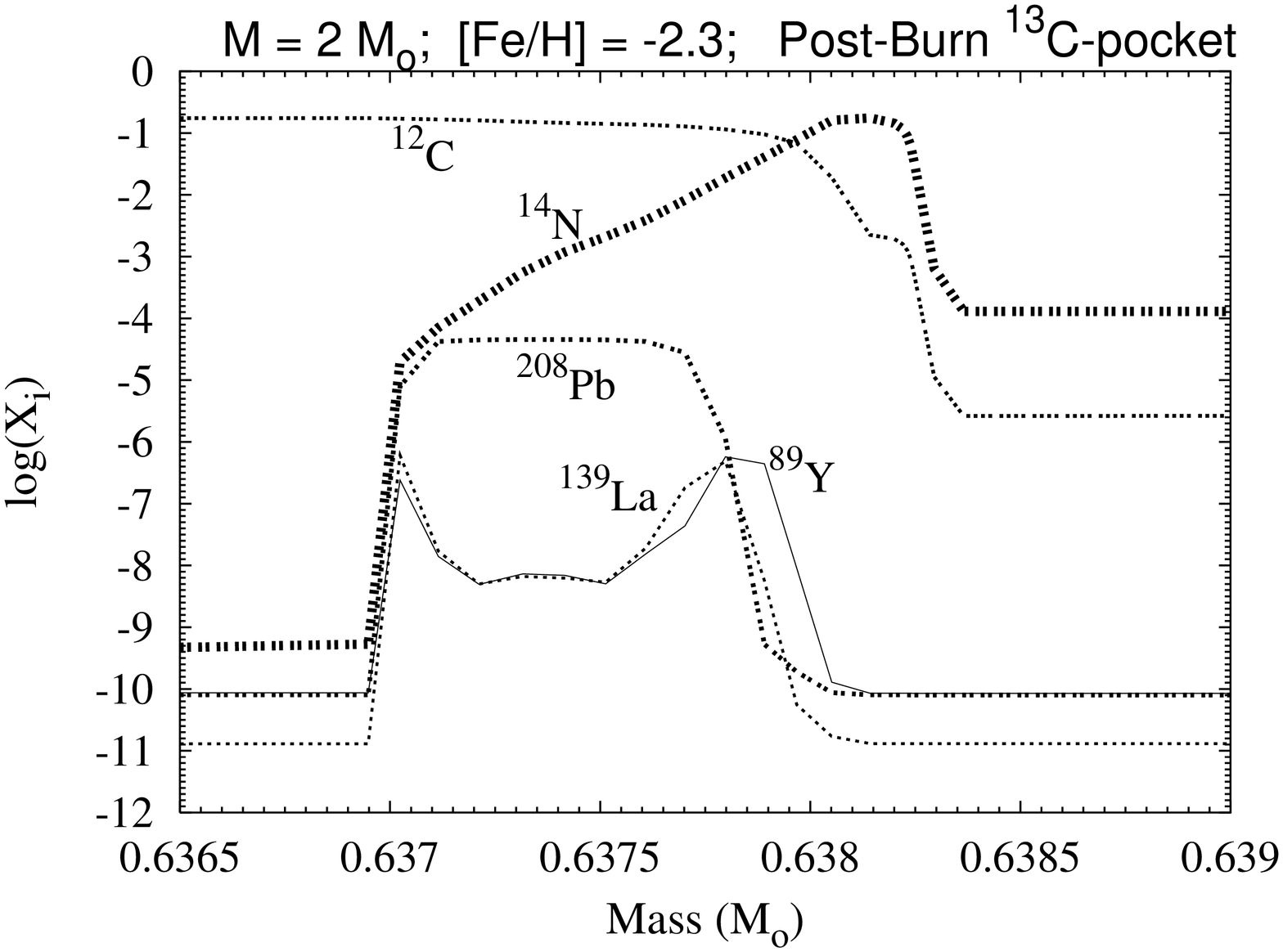}}
\caption{\footnotesize $^{13}$C-pocket mass region for a full
evolutionary AGB model of $M$ = 2 $\Msun$ and [Fe/H] = $-$2.3
(C09) after the first TDU, at the pocket formation
(\textsl{top panel}) and at the end of the $^{13}$C burning
(\textit{bottom panel}).}
\label{sergiopocket}
\end{figure}

C09 introduce a mixing algorithm depending on a free parameter\footnote{See 
C09 for the procedure followed to calibrate it.}
in their full evolutionary models
to mimics the formation of a transition zone
between the fully convective envelope and the radiatively stable
H-exhausted core. Thus, a partial mixing of protons takes place
leading to the formation of a $^{13}$C rich layer. 
Its mass and profile decrease with the number of pulses (see C09,
their Fig.s~4 and~8).
Fig.~\ref{sergiopocket},
top panel, shows this region. In the uppermost layers of the
pocket, where protons are more abundant, the $^{13}$C-pocket is
overlapped with a $^{14}$N-pocket, which forms via the $^{13}$C(p,
$\gamma$)$^{14}$N reaction. $^{14}$N acts as a neutron poison via 
the resonant reaction $^{14}$N(n, p)$^{14}$C, thus subtracting
neutrons to the nucleosynthesis of the $s$-process
elements.
Fig.~\ref{sergiopocket}, bottom panel, shows the same mass region
at the end of the $^{13}$C burning. We concentrate on three
isotopes, $^{89}$Y, $^{139}$La and $^{208}$Pb, chosen as
representative of the three $s$-process peaks. As expected, at
this low metallicity (Gallino et al. 1998), a large amount of
$^{208}$Pb is produced. Maximum Pb production occurs in the
central layers of the pocket, where $X$($^{13}$C) $>$ $X$($^{14}$N)
(we find $X$($^{208}$Pb) = 4.5 $\times$ 10$^{-5}$), while Y
and La show definitely lower abundances: $X$($^{89}$Y) $\sim$
$X$($^{139}$La) $\sim$ 6 $\times$ 10$^{-9}$. In the outer and inner
regions of the pocket, however, $^{89}$Y and $^{139}$La show
peaked distributions. Note that, in the outer tail, $s$-process
elements are efficiently synthesised even if $X$($^{13}$C) $<$
$X$($^{14}$N).
\\
In order to test the effect of these tails on Y, La and Pb with 
different H profiles, we use the post-process nucleosynthesis 
models described by Bisterzo et al. (2010). 
We adopt the H profile of Gallino et al. (1998; case ST; their 
Fig. 2). Then, we introduce a further region in the pocket 
(with mass $M$ = 4 $\times$ 10$^{-4}$ $M_\odot$) where
we change the abundances of $^{13}$C and $^{14}$N to simulate 
different H profiles in the tails.
Therefore, we multiply or divide by different factors the
$^{13}$C and $^{14}$N abundances in the pocket\footnote{In fact,
a range of $^{13}$C-pockets is introduced in order to interpret the
spread in the $s$-elements observed in CEMP-$s$ stars.}.
Note that the H profile and the mass of the pocket are kept 
constant pulse by pulse. The envelope abundances of
the two \spro\ indexes [La/Y] and [Pb/La] obtained with the 
post-process method are show in
Tables~\ref{diff13Cpockets} and~\ref{diffIVzone}.
In Table~\ref{diff13Cpockets}, first group, we show the results
computed with standard $^{13}$C-pockets (i.e. with 3 zones as in 
Gallino et al. 1998) for various
$^{13}$C-pocket efficiencies (from ST $\times$ 2 down to ST/24).
These results are compared with models with an added zone 4 with
$X$($^{13}$C) $<$ $X$($^{14}$N) (Table~\ref{diff13Cpockets}, second
group). This has been done to simulate the effect induced by the
outer tail of the pocket shown in Fig.~\ref{sergiopocket} by C09
model on post-process calculations results. 
C09 obtain a final [La/Y]
= 0.45 and [Pb/La] = 1.30. With the post-process method and a
range of standard $^{13}$C-pockets, [La/Y] reaches a maximum value
of $\sim$ 0.9 (case ST/6) and [Pb/La] $\sim$ 2.2  (case ST/1.5).
When adding the zone 4, minimal variations are found for large and
low $^{13}$C-pocket efficiencies, while appreciable differences
are found in the intermediate cases. For large $^{13}$C abundances
(case ST $\times$ 2), the addition of zone 4 leads to a large 
production of light elements (Ne, Na, Mg), whose poisoning effect
induces a slightly decrease of the final $s$-process element surface
overabundances.
For very low $^{13}$C efficiencies the $s$-process production
is mainly due to the $^{22}$Ne($\alpha$, n)$^{25}$Mg reaction
(Bisterzo et al. 2010),
minimizing the effects of additional $^{13}$C and$^{14}$N.
For intermediate cases, instead, the
introduction of the zone 4 reduces the maximum [La/Y] to $\sim$
0.5 dex and the maximum [Pb/La] to $\sim$ 1.6 dex.
In Table~\ref{diffIVzone} we select the highest $^{13}$C-pocket
case (case ST $\times$ 2) and we test the effect of an added zone 4 with
different $X$($^{13}$C) values ($X$($^{14}$N) is assumed to be negligible).
We choose the ST $\times$ 2 case because previous comparisons done at
larger metallicities (C09) indicate that the
best agreement between post-process and full evolutionary models
is found with this case. The standard case with 3 zones only (II
column) gives [La/Y] = 0.50 and [Pb/La] = 2.04, while the addition
of a zone 4 with $X$($^{13}$C) = 3.8E-4 (test III) definitely
lowers the [Pb/La] ratio (1.47) leaving practically untouched the
[La/Y] ratio (0.56).
Thus, a reasonable agreement between this test and C09 is found even at
such low metallicities.
Verified that the tails of the $^{13}$C-pocket affect the $s$
distribution, one may constrain the choice of the H profile
through a study of spectroscopic observations in CEMP-$s$ stars.
Note that, for disk metallicities, the tails of the pocket do 
not influence sensibly the $s$ distribution.

\begin{table*}
\caption{Envelope abundances of [Y/Fe], [La/Fe], [Pb/Fe]
and their ratios [La/Y] and [Pb/La] for a post-process model
of $M$ = 2 $\Msun$ and [Fe/H] = $-$2.3 and various $^{13}$C-pocket
efficiencies (from ST $\times$ 2 down to ST/24).
The first group lists the results obtained with the standard
$^{13}$C-pocket, 
while in the
second group a further zone 4 with $X$($^{13}$C) $<$
$X$($^{14}$N) is added.}
\label{diff13Cpockets}
\begin{center}
\begin{tabular}{llccccccc}
\hline
Cases &  &  ST$\times$2&   ST &  ST/1.5& ST/2&   ST/6&   ST/12&  ST/24\\[0.5ex]
 \hline
&{[Y/Fe]}   &   1.68& 1.39 & 1.33&   1.35&   1.98&   2.35&   2.45\\
&{[La/Fe]}  &   2.18& 1.88 & 1.92&   2.10&   2.85&   2.94&   2.71\\
&{[Pb/Fe]}   &   4.22&4.12 &  4.09&   4.06&   3.82&   3.44&   2.69\\
&{[La/Y]}   &   0.50&0.49  & 0.59&   0.75&   0.87&   0.59&   0.26\\
&{[Pb/La]}  &   2.04& 2.24 & 2.17&   1.96&   0.97&   0.50&   -0.02\\[1.0ex]
 \hline
 &  &    &   &   & & &   \\
                 & $X$($^{13}$C)= &7.2E-2&3.7E-2 & 2.5E-2&1.9E-2&6.2E-3&3.1E-3&1.6E-3\\
                 & $X$($^{14}$N)= &2.7E-1&1.4E-1 & 9.3E-2&7.1E-2&2.3E-2&1.2E-2&5.8E-3\\[1.0ex]
 zone 4 &{[Y/Fe]}     &    1.58&1.75  & 1.89&   2.00&   2.40&   2.58&   2.64\\
 $X$($^{13}$C) $<$ $X$($^{14}$N)  &{[La/Fe]}    &    2.10& 2.25 & 2.42&   2.54&   2.92&   2.97&   2.80\\
   &{[Pb/Fe]}    &    4.04&4.02  & 3.99&   3.97&   3.76&   3.43&   2.84\\
   &{[La/Y]}     &    0.52&  0.50& 0.53&   0.54&   0.52&   0.39&   0.16\\
                 &{[Pb/La]}    &    1.94& 1.77 & 1.57&   1.43&   0.84&   0.46&   0.04\\
\hline
\end{tabular}
\end{center}
\end{table*}

\begin{table*}
\caption{The same as Table~\ref{diff13Cpockets}, but for a
case ST $\times$ 2 and an added zone 4 with different $X$($^{13}$C) values,
from 0 (standard case) up to 4.3E-3. $X$($^{14}$N) is assumed to be negligible.
In the last column the results obtained by C09 are listed.}
\label{diffIVzone}
\begin{center}
\begin{tabular}{|l|c|ccccccc|c|}
\hline
zone 4,  &standard&  I test& II test&    III test&   IV test&    V test& VI test & VII test &  C09   \\[0.5ex]
 $X$($^{13}$C)  &  0.0 & 2.9E-4& 3.5E-4 & 3.8E-4 & 4.8E-4& 5.8E-4&        1.2E-3&     4.3E-3&     \\[0.5ex]
\hline
{[Y/Fe]}  &     1.68&   2.23 &2.21 & 2.19  &  2.09  &     1.97&           1.74&           1.75&  1.12   \\
{[La/Fe]}   &   2.18&   2.60 &2.65 & 2.75  &  2.80  &        2.76&           2.48&           2.27& 1.57 \\
{[Pb/Fe]}   &   4.22&   4.21 &4.23 & 4.22  &  4.23  &        4.24&           4.29&           4.33& 2.88 \\
{[La/Y]}    &   0.50&   0.37 &0.44 & 0.56  &  0.71  &        0.79&           0.74&           0.52& 0.45 \\
{[Pb/La]}   &   2.04&   1.61 &1.58 & 1.47  &  1.43  &        1.48&           1.81&           2.06& 1.30 \\
\hline
\end{tabular}
\end{center}
\end{table*}

\section{Conclusions}

The maximum amount of $^{13}$C and $^{14}$N in the pocket and
different hydrogen profiles (and therefore the amount of $^{13}$C
and $^{14}$N in the tails of the pocket) modify the s abundance
distribution. In particular, the $s$-process indexes [La/Y] and
[Pb/La] are sensitive to the tails of the pocket. 
At [Fe/H] = $-$2.3, a large amount of $^{208}$Pb is produced when $X$($^{13}$C)
$>$ $X$($^{14}$N). A first interesting consequence caused by the
addition of an outer tail in the pocket with $X$($^{13}$C)
$<$ $X$($^{14}$N) is that the maximum [La/Y]
value attained with different $^{13}$C-efficiencies is reduced to
$\sim$ 0.5. Moreover, when assuming a calibrated extra
$X$($^{13}$C) in the tails of the pocket, the maximum
[Pb/La] is reduced to 1.4 dex.
Comparison between theory and observations in CEMP-$s$ stars
are then needed in order to constrain the choice of the H profile 
in the central and outer regions of the $^{13}$C-pocket during AGB 
nucleosynthesis.

\textit{Acknowledgements}\\
We are grateful to Roberto Gallino for helpful comments and discussions.

\bibliographystyle{aa}

\end{document}